\documentclass[useAMS,usenatbib]{mn2e}
\usepackage{graphicx}

\title[Shock oscillation model for QPOs in  black holes]
{Shock oscillation model for QPOs in stellar-mass and supermassive black holes}
\author[T. Okuda, V. Teresi  and D. Molteni]{T. Okuda$^{1}$
\thanks{E-mail:okuda@cc.hokkyodai.ac.jp}, V. Teresi$^{2}$,  
D. Molteni$^{2}$ \\
$^{1}$Hakodate Campus, Hokkaido University of Education, Hachiman-cho 1-2,
 Hakodate 040-8567, Japan\\
$^{2}$Dipartimento di Fisica e Tecnologie Relative, Universita di Palermo,
 Viale delle Scienza,  Palermo, 90128, Italy}
 
\begin{document}

\date{Accepted }

\pagerange{\pageref{firstpage}--\pageref{lastpage}} \pubyear{2004}

\maketitle

\label{firstpage}

\begin{abstract}
 We  numerically examine centrifugally supported shock waves in 2D rotating 
 accretion flows  around a stellar-mass ($10 M_{\odot}$) and a supermassive 
 ($10^6 M_{\odot}$) black holes over a  wide range of input accretion rates of 
 $10^7 \ge \dot M/\dot M_{\rm E} \ge 10^{-4}$.
 The resultant 2D-shocks are unstable with time and the luminosities show 
 quasi-periodic oscillations (QPOs) with modulations of a factor of 2 -- 3 and 
with periods of a tenth seconds to several hours, depending on the black hole masses.
 The shock oscillation model may explain the intermediate frequency QPOs with 1-10 Hz 
 observed in the stellar-mass black hole candidates 
 and also suggest the existence of QPOs with the period of hours in AGNs.
 When the accretion rate $\dot M$ is low, the luminosity increases in proportion 
to the accretion rate. However, when $\dot M$  exceeds greatly the Eddington critical
 rate $\dot M_{\rm E}$, the luminosity is insensitive to the accretion rate and 
 is kept constantly around $\sim 3 L_{\rm E}$.
 On the other hand, the mass-outflow rate $\dot M_{\rm loss}$ increases in proportion to 
 $\dot M$ and it amounts to about a few percent of the input mass-flow rate.

\end{abstract}


\section{Introduction}
 Rotating inviscid and  adiabatic accretion flow around a black hole 
 can have two saddle type sonic points.
 After the flow with adquate injection parameters passes through 
 the outer sonic point, the centrifugal force can virtually stop the rotating 
 supersonic flow, forming a standing shock close to
 the black hole.  Then, the flow passes through the inner sonic point and again falls
into the black hole supersonically.
 In such transonic problems of accretion and wind,
 \citet{b11} and \citet{b1}  firstly showed satisfactory analytical or 
 numerical global solutions under a full relativistic treatment 
 and under a pseudo-Newtonian potential, respectively.
 It has been shown that these generalized accretion flows could be responsible
 for the hard and soft state transitions or the quasi-periodic oscillations
 (QPOs) of the hard X-rays from the black hole candidates 
 \citep{b19,b29,b15}.
  Further analyses of the transonic problems under modified  pseudo-Newtonian potentials
 showed that the standing shocks are essential ingredients in  multi-transonic black hole
 accretion discs \citep{b8,b6,b7} and that the generalized multi-transonic accretion model
 may show  QPO frequencies of Galactic black hole candidates in terms of dynamical flow 
 variables \citep{b9,b10}.

 Recently, in relative to the shock oscillation models of QPOs, 
 \citet{b26} and \citet{b3} showed 
 several numerical simulations of 2D accretion flows around black holes,
 using the Eulerian method under the treatment of radiation transport and the SPH method
 in the presence of cooling effects, respectively. 
 From the power  spectra of the luminosities, they showed that QPOs 
 are found at a few Hz to hundreds of Hz for stellar-mass
 black holes with mass of $10 M_{\rm \odot}$ and oscillation time-scales of hours to weeks 
 for supermassive black holes with mass of $10^8 M_{\rm \odot}$.
In the present paper, we examine  further shock oscillation models over a wide range 
of accretion rates, focusing on the general properties of the shock waves
and the QPO frequencies  in the 2D nonadiabatic accretion flows around the black holes.

\section{ Model Equations}
  A set of relevant equations consists of six partial differential 
 equations for density, momentum, and thermal and radiation energy.
 These equations include the heating and cooling of gas and 
 radiation transport.   
 The radiation transport is treated in the gray, flux-limited diffusion
 approximation \citep{b16}.
 We  use spherical polar coordinates ($r$,$\zeta$,$\varphi$), where $r$
 is the radial distance, $\zeta$ is the polar angle measured from 
 the equatorial plane of the disc, and $\varphi$ is the azimuthal angle.  
 The gas flow is assumed to be axisymmetric with respect to $Z$-axis 
 ($\partial /\partial \varphi=0 $) and the equatorial plane .
 In this coordinate system, we have the basic
 equations in the following conservative form \citep{b14}:

%%%%%%%%  equation (1)     %%%%%%%%%%%%%%%%%%%%%%%%%%%%%%%%%%%%%%
 \begin{equation}
   { \partial\rho\over\partial t} + {\rm div}(\rho\bmath{v}) =  0,  
 \end{equation}
%%%%%%%  equation (2) %%%%%%%%%%%%%%%%%%%%%%%%%%%%%%%%%%%%%%%%%%
 \begin{equation}   
  {\partial(\rho v)\over \partial t} +{\rm div}(\rho v \bmath{v})  =
  \rho\left[{w^2\over r} + {v_\varphi^2\over r}-{GM_* \over (r-r_{\rm g})^2} 
  \right] -{\partial p\over \partial r}+f_r,
\end{equation}
%%%%%%%  equation (3) %%%%%%%%%%%%%%%%%%%%%%%%%%%%%%%%%%%%%%%%%%
 \begin{equation}  
  {{\partial(\rho rw)}\over \partial t} +{\rm div}(\rho rw\bmath{v}) 
  = -\rho v_\varphi^2{\rm tan}\zeta-{\partial p\over\partial\zeta}
    + f_\zeta, 
 \end{equation}
%%%%%%%  equation (4)  %%%%%%%%%%%%%%%%%%%%%%%%%%%%%%%%%%%%%%%
 \begin{equation}    
 {{\partial(\rho r{\rm cos}\zeta v_\varphi)}\over \partial t} 
     +{\rm div}(\rho r{\rm cos}
 \zeta v_\varphi\bmath{v}) = 0, 
\end{equation}
%%%%%%%  equation (5)  %%%%%%%%%%%%%%%%%%%%%%%%%%%%%%%%%%%%%
\begin{equation}  
  {{\partial \rho\varepsilon}\over \partial t}+
    {\rm div}(\rho\varepsilon\bmath{v})
      = -p\;\rm div \bmath{v}  - \Lambda, 
\end{equation}
%%%%%%%  equation (6)  %%%%%%%%%%%%%%%%%%%%%%%%%%%%%%%%%
and
\begin{equation}       
  {{\partial E_0}\over \partial t}+ {\rm div} \bmath{F}_0+
        {\rm div}(\bmath{v}E_0 + \bmath{v}\cdot P_0) 
        = \Lambda 
      - \rho{(\kappa +\sigma)\over c}\bmath{v}\cdot
      \bmath{F}_0,
 \end{equation} 
 where $\rho$ is the density, $\bmath{v} =(v, w, v_\varphi)$ are the
 three velocity components, $G$ is the gravitational constant,
 $M_*$ is the central mass, $p$ is the  gas pressure,
 $\varepsilon$ is the specific internal energy of the gas,  $E_0$ is 
 the radiation energy density per unit volume, $P_0$ is the radiative
  stress tensor, and $c$ is the speed of light. The subscript ``0'' denotes
  the value in the comoving frame and that the equations are correct
   to the first order of $\bmath{v}/c$ \citep{b13}.
 We adopt the pseudo-Newtonian potential \citep{b27}
 in equation (2), where $r_{\rm g}$ is the Schwarzschild radius.
 The force density $\bmath{f}_{\rm R}=(f_r,f_\zeta)$ exerted 
 by the radiation field is given by
%%%%%%% equation  (7)    %%%%%%%%%%%%%%%%%%%%%%%%%%%%%%%%%%%%%%%%%%%%%% 
\begin{equation} 
  \bmath{f}_{\rm R}=\rho\frac{\kappa+\sigma}{c}\bmath{F}_0, 
 \end{equation} 
 where $\kappa$ and $\sigma$ denote the absorption and scattering 
 coefficients and $\bmath{F}_0$ is the radiative flux 
 in the comoving frame.
 The quantity $\Lambda$ describes the cooling and heating of the gas, 
 
%%%%%%% equation (8) %%%%%%%%%%%%%%%%%%%%%%%%%%%%%%%%%%%%%%%%%%%
 \begin{equation}      
      \Lambda = \rho c \kappa(S_*-E_0), 
\end{equation}
 where $S_*$ is the source function. 
 For this source function, we assume local thermal equilibrium $S_*=aT^4$, 
 where $T$ is 
 the gas temperature and $a$ is the radiation constant.
 For the equation of state, the gas pressure is given by the ideal gas law, 
 $p=R_{\rm G}\rho T/\mu$, where $\mu$ is the mean molecular weight 
 and $R_{\rm G}$ is the gas constant. 
  The temperature $T$ is proportional to the specific
 internal energy, $\varepsilon$, by the relation $p=(\gamma-1)\rho\varepsilon
  =R_{\rm G}\rho T/\mu$, where $\gamma$ is the specific heat ratio.  
  To close the system of 
 equations, we use the flux-limited diffusion approximation  
 for the radiative flux:
%%%%%%% equation (9)  %%%%%%%%%%%%%%%%%%%%%%%%%%%%%%%%%%%%%%%%%%%
\begin{equation}
   \bmath{F}_0= -{\lambda c\over \rho(\kappa+\sigma)}
   {\rm grad}\;E_0, 
\end{equation}

\noindent and
%%%% equation (10) %%%%%%%%%%
\begin{equation}
   P_0 = E_0 \cdot T_{\rm Edd}, 
\end{equation}
where  $\lambda$ and $T_{\rm Edd}$ are the {\it flux-limiter} and the 
 {\it Eddington Tensor}, respectively, for which we use the approximate
 formulas given in \citet{b14}.
 The formula fulfill the correct limiting conditions $\lambda\rightarrow
  1/3 $ in the optically thick diffusion limit,
  and $\mid\bmath{F}\mid\rightarrow cE_0 $ as $\lambda\rightarrow 0$ 
  for the optically thin streaming limit.

\section{Numerical Methods}

 The set of partial differential equations (1)--(6) is 
 numerically solved by a finite-difference method under adequate initial 
 and boundary conditions.
 The numerical schemes used are basically the same as that described  
  previously \citep{b14,b25}. 
 The methods are based on an explicit-implicit finite difference scheme.

 \subsection{Model Parameters}
 We consider a stellar-mass black hole with mass $M_*= 10 M_{\odot}$ and a supermassive
 black hole with $M_*= 10^6 M_{\odot}$.
 To examine the shock model, we  determine the injection parameters, such as 
 the specific angular momentum, $\lambda_{\rm out}$, the radial velocity, $v_{\rm out}$, 
 and the sound velocity, $a_{\rm out}$, at an outer boundary radius, $R_{\rm out}$, whose 
  parameters can lead to  a  shock wave close to the black hole.
  We search analytically the injection parameters through an examination 
  of the parameter space ($E_{\rm out}, \lambda_{\rm out}$), where $E_{\rm out}$
  is the total specific energy\citep{b1,b18,b20}.
  The typical model parameters used
  are listed in Table 1. Here, the velocities and the distances are given in units of 
  $c$ and $r_{\rm g}$, respectively, and 
  $\dot m$ is the input accretion rate normalized to the Eddington critical accretion rate
  $\dot M_{\rm E}( =  L_{\rm E} / c^2$),
 where $L_{\rm E}$ is the Eddington luminosity given by
 
 \begin{equation}
   L_{\rm E}  = {4\pi GM_* c \over {\kappa_{\rm e} }},
 \end{equation}
 and  $\kappa_{\rm e}$ is the electron scattering opacity.
 $L_{\rm E}$ and $\dot M_{\rm E}$ are 
 1.5 $\times 10^{39}$ erg $s^{-1}$ and $1.7 \times 10^{18}$ g $s^{-1}$,
 respectively, for the stellar-mass black hole with $M_*=10 M_{\odot}$.
 $\phi$ is the subtended angle of the central black hole to the disc
 height $h$ at $r=R_{\rm out}$, that is, ${\rm tan} \phi=(h/r)_{\rm out}$.
 The inner-boundary radius $R_{\rm in}$ of the computational domain
 is taken to be  $2 r_{\rm g}$.

\begin{table*}
\centering
\caption{Model parameters}
\begin{tabular}{@{}cccccccc} \hline \hline
 $M_*/M_{\odot}$ & $\lambda_{\rm out}$ & $v_{\rm out}$ & $a_{\rm out}$ & 
  $\rho_{\rm out}$($\rm g \;cm^{-3}$) & $\dot m$ & $\phi$ & $
        R_{\rm out}$ \\\hline
  $10 $  & 1.64 & 0.0751   & 0.0738    & 
  $2.1 \times 10^{-12}$ -- $7 \times 10^{-2}$ &  $10^{-4}$ -- $7\times 10^6$ &
  $ 29^{\circ}$  & 30 \\ $10^6$ & 1.875  & 0.0751   & 0.0654       & 
   $2.1 \times 10^{-17}$ -- $10^{-6}$  & $10^{-4}$ -- $ 10^7$ & $ 29^{\circ}$  & 30 
   \\ \hline
\end{tabular}
\end{table*}

\subsection{ Initial and Boundary Conditions}
  At the outer-disc boundary we assume a continuous inflow of matter 
 with a constant accretion rate $\dot M$ and the injection flow parameters
 in Table 1.  
 For the accretion rates considered here, we have an optically thick and 
 radiation-pressure dominant accretion flow at the outer boundary $R_{\rm out}
  = 30r_{\rm g}$.
 Then, the input gas temperature $T_{\rm out}$ is given by 
 
   \begin{equation}
    (a_{\rm out})^2 \sim {4\over 3}({P\over \rho})_{\rm out} 
    \sim  {4\over 9} ({a T^4 \over \rho})_{\rm out},
   \end{equation}
 where $P$ is the total pressure.
 Therefore, for a given $a_{\rm out}$, we have a smaller $T_{\rm out}$ for
   a smaller ambient density $\rho_{\rm out}$.
 The initial conditions  of the flow except the outer-disc 
 boundary are adequately given as a radially hydrostatic equilibrium state
 with zero azimuthal velocities everywhere.
 Physical variables at the inner boundary $R_{\rm in}=2r_{\rm g}$, 
 except for the velocities,
 are given by extrapolation of the variables near the boundary. 
  However, we impose limited conditions that the radial velocities are given 
  by a free-fall velocity and the angular velocities are zero.
  On the rotational axis and the equatorial plane, 
 the meridional tangential velocity $w$ 
 is zero and all scalar variables must be symmetric relative to these axes.
 As to the outer boundary region above the outer-disc, we  use free-floating 
 conditions  and allow for outflow of matter, whereas any 
 inflow is prohibited here.
  With these initial and boundary conditions, we perform  time 
 integration of equations (1)--(6) until a quasi-steady solution is obtained. 
 
       \begin{figure}
       \includegraphics[width=86mm,height=66mm]{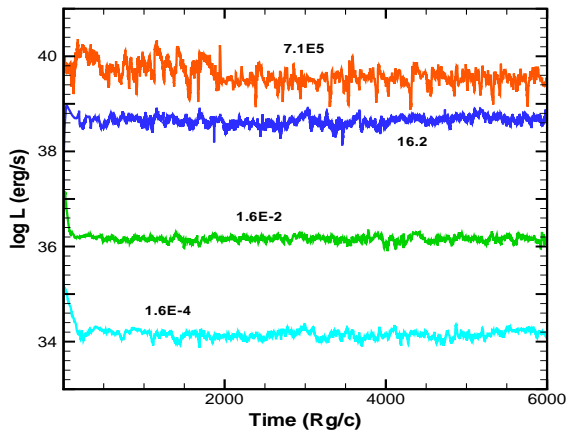}
       \caption{Time evolutions of luminosity $L$ (erg s$^{-1}$)  
       for $\dot m = 1.6\times 10^{-4}, 1.6 \times 10^{-2}$, 
          16.2, and $ 7.1 \times 10^5$ for the stellar-mass black hole. }
      \label{fig1}
      \end{figure}

 \begin{figure}
      \includegraphics[width=86mm,height=66mm]{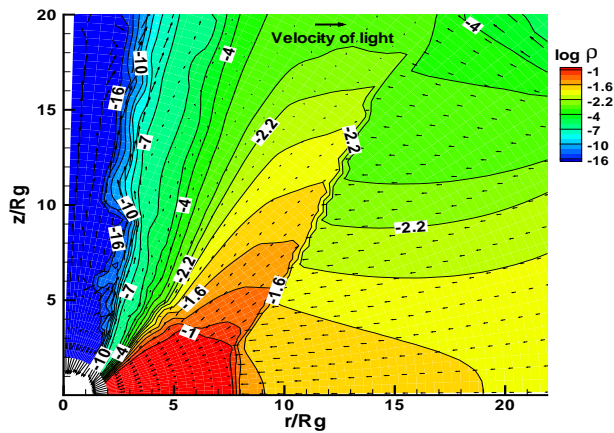}
      \caption{Density contours and velocity vectors at $t = 1.2\times 
      10^4 \;(r_{\rm g}/c)$ for $\dot m = 7.1\times 10^5$ of the stellar-mass
       black hole.
       The standing shock is formed at $r/r_{\rm g} \sim 8$ and  
       $z/r_{\rm g} \leq 3$ near to the equatorial plane but it
        is bent obliquely toward upstream.}  
      \label{fig2}
      \end{figure}

 \section{Numerical Results}

 \subsection{Stellar-Mass Black Hole}

      \begin{figure}
      \includegraphics[width=86mm,height=66mm]{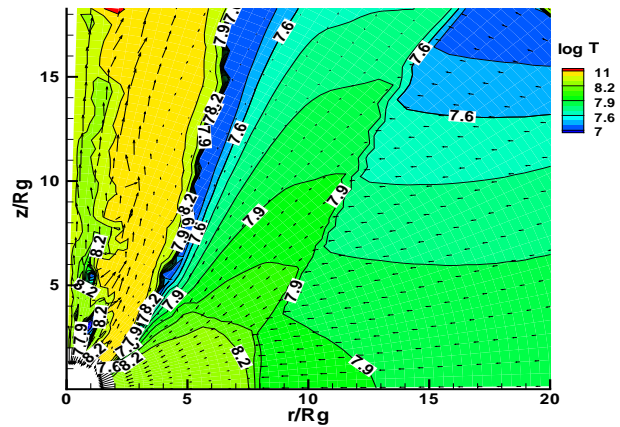}
      \caption{Same as Fig.~2 but temperature contours.
       The black thick line behind the shocked region shows the funnel wall
       which is characterized by vanishing effective potential.}  
      \label{fig3}
      \end{figure}
 
     \begin{figure}
       \includegraphics[width=86mm,height=66mm]{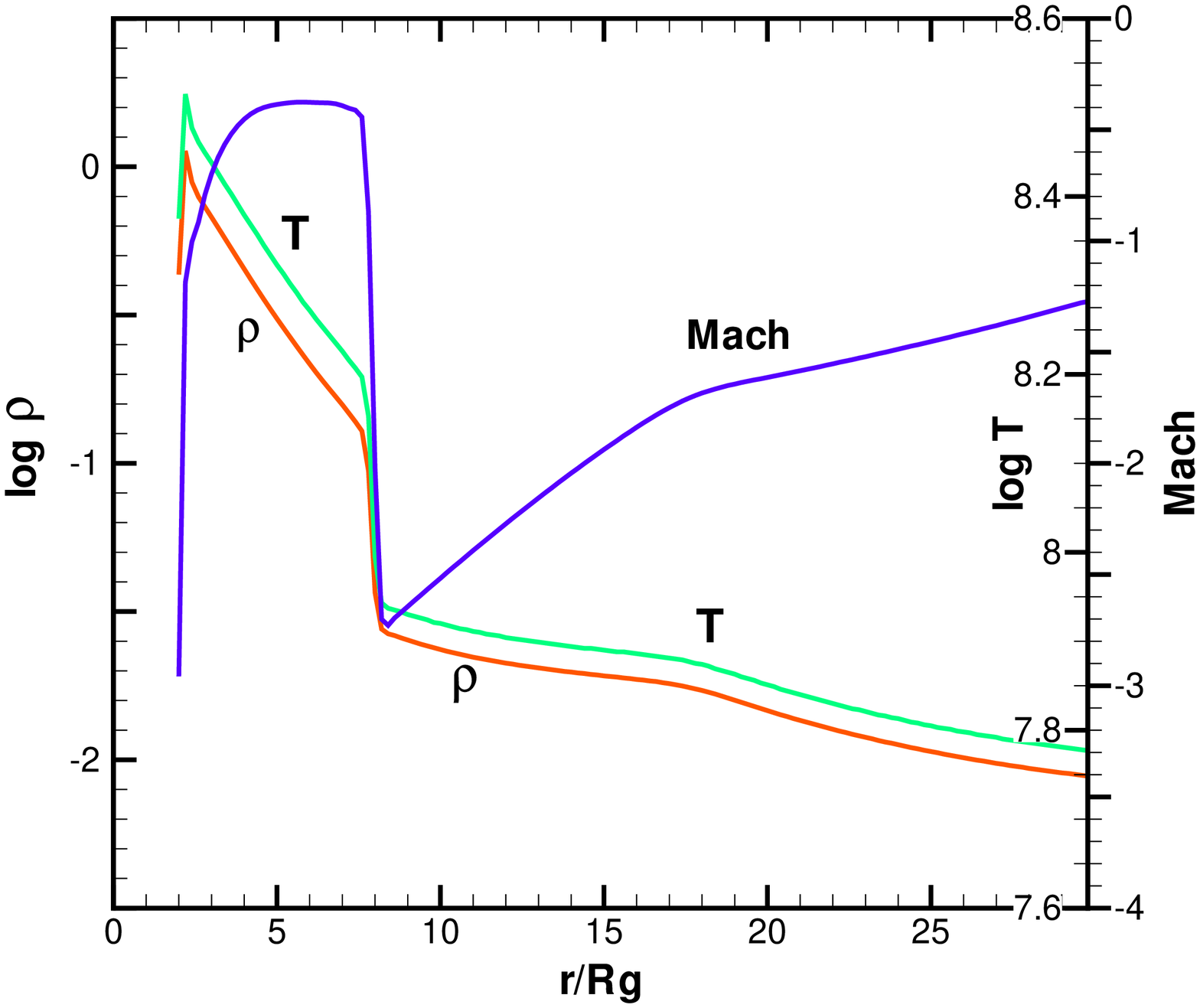}
       \caption{Profiles of density $\rho$ (g cm$^{-3}$),  temperature $T$ (K), and  Mach number 
       of the radial velocity on the equatorial plane at 
       $t = 1.2\times 10^4\; (r_{\rm g}/c)$ for 
       $\dot m=7.1 \times 10^5$ of the
       stellar-mass black hole.}
      \label{fig4}
      \end{figure}

  The input flow with the injection parameters at the outer-disc radius
  $R_{\rm out}$  arrives at the innermost radius in a time smaller than 
   $R_{\rm out}/v_{\rm out}(= 400 r_{\rm g}/c)$ and then the system is settled down
   toward a quasi-steady state configuration. 
   The luminosity curve is a good measure to check if the quasi-steady state 
   is attained.
  Fig.~1 shows the time evolutions of luminosity $L$ (erg s$^{-1}$) 
  for casese of $\dot m = 1.6\times 10^{-4}, 1.6 \times 10^{-2}$, 
  16.2, and $ 7.1 \times 10^5$ for the stellar-mass black hole. 
  In each case a quasi-steady state is obtained and the centrifugally 
   supported shock is formed around the black hole. 
   However, the luminosities show QPO phenomena with modulations 
   of a  factor of 2 -- 3
  and the shock positions on the equatorial plane are also found to be 
  variable around several Schwarzschild radii. 
  
  The overall features of the flow and the shock  are shown in Figs.~2 and 3
  which show the contours of density $\rho$ (g cm$^{-3}$) and temperature $T$ (K) with 
  velocity vectors at the evolutionary time $t= 1.2\times 10^4 (r_{\rm g}/c)$ 
  for  $\dot m =7.1\times 10^5$, respectively. 
  Here, a spheroidal shock surface is found, where
  a standing shock exists at $r/r_{\rm g} \sim 8$ and $z/r_{\rm g} \le 3 $ 
  near to the equatorial plane  but it is bent obliquely toward upstream. 
 The densities and the temperatures are enhanced by a several factor 
  across the shock surface. Behind the shocked region there exists a 
  funnel wall which is characterized by vanishing effective potential and is
  roughly denoted by the black thick line in Fig.~3. 
  The extended shocked region between the shock surface and the funnel wall consists
  of the hot and dense gas with decelerated velocities.
  The luminosity modulations are caused by oscillations of the hot shocked region.
  In the cone-like funnel region between the rotational axis and the funnel wall,
  the temperatures are much higher but the densities are very low,
   and the gas is optically thin.  The accreting matter in the inner shocked region
  mostly flows into the black hole and partly diverts into the funnel region.
  The outflow gas generated there is originally subsonic but is accelerated to
  relativistic velocities due to the radiation pressure force.
  An upper arrow in Fig.~2 shows the reference velocity vector of light.
  
  Fig.~4 denotes the shock profiles of density $\rho$ (g cm$^{-3}$), 
  temperature $T$ (K), and  Mach number 
  of the radial velocity on the equatorial plane at the same evolutionary time
  as Figs.~2 and 3.
  The radially infalling gas attains to Mach number of $\sim 2.7 $ at the shock 
  and is abruptly decelerated across the shock front, and is again 
  supersonically swallowed into the black hole.
  To understand the transitions of flow variables at the
    shock, we refer to the pressure balance equation of
   Rankine--Hugoniot relations:
   \begin{equation}
     P_2 + \rho_2 {v_2}^2 = P_1 + \rho_1 {v_1}^2,
   \end{equation}
   where $P$ is the total pressure due to radiation pressure $f_{\rm E} E_0$ and gas 
   pressure $R_{\rm g}\rho T/\mu$, $f_{\rm E}$ is the Eddington factor \citep{b14}, 
   and  subscripts ``1'' and ``2'' describe quantities before and after the shock.

  In the case of $\dot m= 7.1\times 10^5$,  the radiation pressure 
  dominates the gas pressure and the gas is optically thick everywhere
   except the funnel region.
  The shock features in Figs.~2--4 are very similar to those in the adiabatic 
  flows \citep{b26}. In Fig.~4, the pre-shock temperature of $\sim 7.5 \times 10^7$ K 
  jumps to the post-shock temperature of $1.5 \times 10^8$ K.
  Here, $f_{\rm E}=1/3, E_0=aT^4, \rho_1 {v_1}^2 \gg P_1$, and $P_2 \gg
  \rho_2 {v_2}^2$.  Then, we have $a{T_2}^4/3 \sim \rho_1{v_1}^2$ from equation (13)
  and $T_2 \sim 1.5 \times 10^8$ K with $\rho_1 = 10^{-1.6}$ g cm$^{-3}$
  and $v_1=0.24$.
  From the numerical data, $\rho_2/\rho_1 = v_1/v_2 \sim 5$. 
  The gas behaves as an adiabatic flow for a perfect gas with $\gamma=1.5$.
  The flux limiter $\lambda$ expresses the degree of optical thickness of 
  the gas and $\lambda$ is 1/3 throughout the whole region  except the funnel region.
  On the other hand, for the low  accretion rates of $\dot m= 1.6 \times 
  10^{-2}$ and $1.6 \times 10^{-4}$,
  the gas is  optically thin everywhere and the shock features are 
  considerably differ from the adiabatic case, and the shock position near 
  to the equatorial plane moves outward compared with those in the high 
   accretion rates.
  
   Fig.~5 denotes the power energy density spectra of luminosity $L$ 
   (red line) and shock position $R_{\rm s}$ (blue line) on the equatorial plane 
  corresponding to Fig.~1. From the power spectra of $L$, we recognize 
  the QPO-frequencies of a few to 10 Hz 
  and the frequencies increase with  increasing $\dot m$. 
  The power spectra of  $R_{\rm s}$  agree qualitatively with that of $L$.

    \begin{figure}
       \includegraphics[width=86mm,height=66mm]{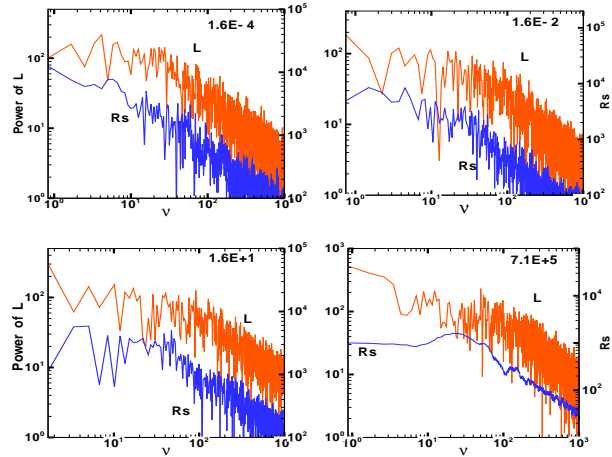}
       \caption{Power energy density spectra of the luminosity $L$ 
       (red line) and the shock position $R_{\rm s}$ (blue line)  
        on the equatorial plane corresponding to Fig.~1. }        
      \label{fig5}
      \end{figure}

       \begin{figure}
       \includegraphics[width=86mm,height=66mm]{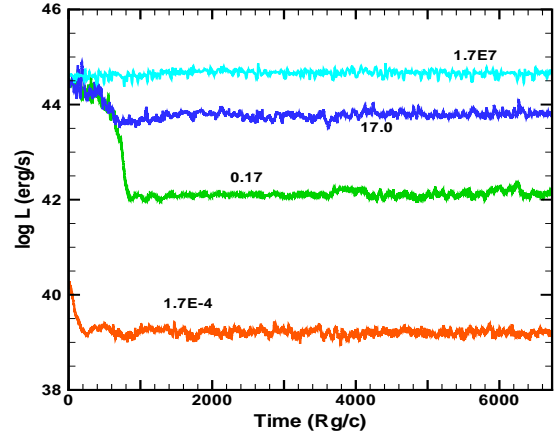}
       \caption{Time evolutions of luminosity $L$ for $\dot m = 1.7
        \times 10^{-4}$, 0.17, 
          17.0, and $ 1.7 \times 10^7$ for the supermassive black hole. }
      \label{fig6}
      \end{figure}

  \begin{figure}
       \includegraphics[width=86mm,height=66mm]{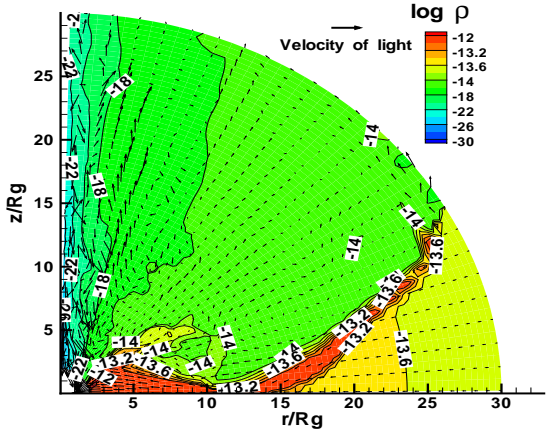}
       \caption{ Density contours and velocity vectors at $t = 7.4\times 10^3
      (r_{\rm g}/c)$
      for $\dot m = 0.17$ of the supermassive black hole.
       The oscillating shock near to the equatorial plane is formed at $r/r_{\rm g}
       \sim$ 12 -- 14 and it extends obliquely toward upstream.}
      \label{fig7}
  \end{figure}
  \begin{figure}
       \includegraphics[width=86mm,height=66mm]{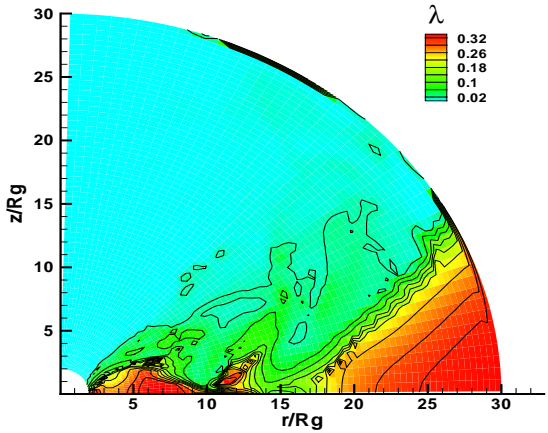}
       \caption{Same as Fig.~7 but the flux-limiter $\lambda$ which 
       shows the degree of optical thickness of the gas.}
      \label{fig8}
  \end{figure}

     \begin{figure}
      \includegraphics[width=86mm,height=66mm]{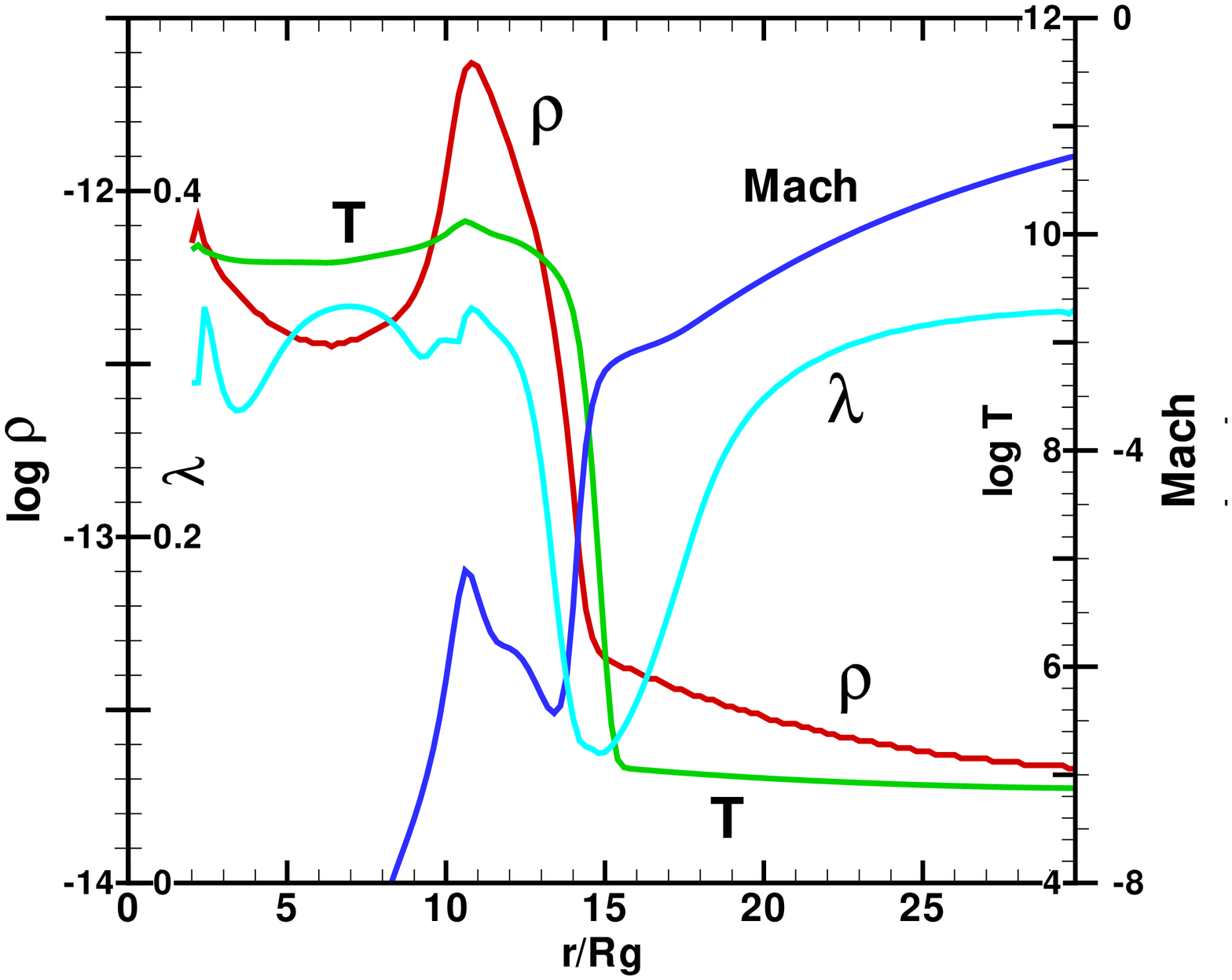}
      \caption{Profiles of  density $\rho$ (g cm$^{-1}$),  
      temperature $T$ (K),   Mach number 
      of the radial velocity, and  flux limiter $\lambda$ 
       on the eaqatorial plane at $t= 7.4 \times 10^3\;(r_{\rm g}/c)$ for 
       $\dot m= 0.17$ of the
       supermassive  black hole.}  
      \label{fig9}
      \end{figure}

 \subsection{Supermassive Black Hole}
   Fig.~6 shows the  luminosity curves 
  for casese of $\dot m = 1.7\times 10^{-4}$, 0.17, 
  17.0, and $ 1.7 \times 10^7$ for the supermassive black hole.
  The luminosities show also the quasi-periodic variations with modulations of 
  a factor of 2 -- 3.
  The absolute luminosities are five orders of magnitude larger than those 
  corresponding to the stellar-mass black hole, because of the much larger 
  mass of the supermassive black hole.
  The general properties of the flows, the shock profiles, 
  and the shock positions for the supermassive black hole are same as those for
   the stellar-mass black hole.  The properties of the flows
   depend on the input accretion rates.
  In the high accretion rates, the shock features are same as Figs.~2 -- 4 for 
  the stellar-mass black hole and the shock
  locates at $r/r_{\rm g} \sim$  7 -- 8 on the equatorial plane, as is 
  thoretically predicted in the adiabatic case.  
  However, in the low accretion rates, the accretion flows differ considerably
   from the cases of  high $\dot m$.

   Figs.~7 and 8 show the contours of density and flux-limiter $\lambda$
   with velocity vectors at $t= 7.4\times 10^3 (r_{\rm g}/c)$ for 
  $\dot m =0.17 $, respectively, where the ambient density $\rho_{\rm out}$ 
  is as low as $\sim 10^{-14}$ g cm$^{-3}$.
   Compared with the flows in Figs.~3 and 4 for the high accretion rate,
   the accretion flow is geometrically  thin in the inner region
  due to the strong radiative cooling effects.
  The hot and dense shocked region is confined in 
   a narrow region (red zone in Fig.~7) above the equatorial plane, 
   where the temperatures are as high as 
   $10^8$ -- $10^{10}$ K. Outside of the shocked region, the densities are
  very low, so that the gas is optically thin and the temperatures are very high
  as well.  
  The shock feature near to the equatorial plane is found at $r/r_{\rm g}
   \sim$ 11 -- 14 and it extends obliquely toward upstream. 
  The optically thick input gas becomes  optically thin ($\lambda \ll 1/3$) 
  in the pre-shock region and is highly compressed at the shock,
  and becomes again optically thick ($\lambda \sim 0.3$)
  in  the post-shock region. 
  
  The profiles of density $\rho$ (g cm$^{-3}$), temperature $T$ (K),  
  Mach number, and 
   flux limiter $\lambda$ on the eqatorial plane are shown in Fig.~9.
  We find here discontinuous structures of the flow variables like a 
  shock wave in the range of $r/r_{\rm g}$ = 11 -- 14  but the flow 
   structures differ considerably from the adiabatic shock solution. 
  Since the upstream temperatures before the discontinuity 
  are as low as $\sim 10^5$ K due to the very low input density,  
  the sound velocity is small and Mach number of the infalling gas becomes large.
  This results in large Mach number of $\sim 6.4$  at the discontinuity.
  Across the discontinuity, the gas is decelerated down to Mach number $\sim $ 5.2 and
  is again supersonically falling into the black hole, while the infalling gas never
   become subsonic after the passage of the outer sonic point and the flow
   can not have two saddle type sonic points.
   The discontinuity found in Fig.~9 is eventually regarded as a pseudo-shock front.
   However, it should be noted that the detailed structures of the horizontal flows 
   at $5 \ge z/r_{\rm g} \ge 1$ above the equatorial plane in Fig.~7
   show  the shock features with the two sonic points and that the shocked 
   regions are joined to the above discontinuity region on the equatorial plane.
   Then we treat the pseudo-shock like a shock wave to estimate the flow variables
   before and after the front. 
   Since the gas pressure dominates the radiation pressure in the post-shock region, 
  we have approximately $R_{\rm g}\rho_2 T_2/\mu = \rho_1 {v_1}^2$. 
  As the result, the low pre-shock temperature of $\sim  10^5$ K   jumps to the 
  very high post-shock temperature $T_2 \sim 10^{10}$ K at the shock.
  The radiation energy density never change so drastically as the temperature
  and the density jumps across the shock and is in same order of $\sim a{T_1}^4$  
  near to the equatorial plane, so that the radiation pressure dominates the gas pressure
  in the pre-shock region and the gas pressure is dominant in the post-shock region.
  The effective thickness of the shock above the equatorial plane  becomes 
  broad instead of the discontinuous one in Fig.~4. This agrees with the previous 
  1D result that the shock thickness broadens with decreasing ambient densities
  \citep{b26}.
  
  Fig.~10 denotes the power  spectra of luminosity $L$ 
   (red line) and shock position $R_{\rm s}$ (blue line) on the equatorial plane 
   corresponding to Fig.~6. 
   From the power spectra of the luminosities, we find the QPO frequencies 
   $\nu_{\rm qpo} = 3\times 10^{-5}$ -- $3\times 
   10^{-4}$ Hz and the period $P_{\rm qpo} = 1$ -- $10$ hours.

  \begin{figure}
       \includegraphics[width=86mm,height=66mm]{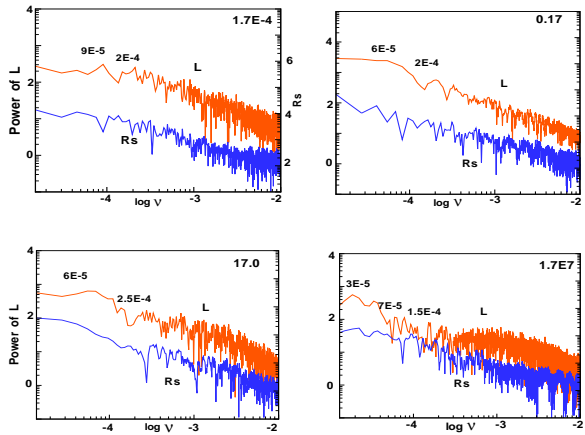}
       \caption{Power energy density spectra of the luminosity $L$ 
       (red line) and the shock position $R_{\rm s}$ (blue line)  
        on the equatorial plane corresponding to Fig.~6}
      \label{fig10}
  \end{figure}

 \section{Shock Location, Luminosity, and Mass-outflow rate}
 As to the shock waves examined here, we find general properties
 of the shock features independent of the black hole masses.  
 This is naturally understood, to some extent, because the 
 injection parameters ($\lambda_{\rm out}, v_{\rm out}, a_{\rm out}$)
 have been derived from the dimensionless
 mass, momentum, and energy conservation equations for the adiabatic gas 
 which are explicitly independent of the black hole masses.
 There are also common properties of normalized  luminosities,  shock
 positions, and  mass-outflow rates versus  accretion
 rate $\dot m$ between the stellar-mass and the supermassive black holes. 
  Fig.~11 shows the shock position $R_{\rm s}$ versus $\dot m$ near to the equatorial 
 plane for the black holes.
 For  $ \dot m \ge 10^3 $, $R_{\rm s}/r_{\rm g}  \sim$  7 -- 8, while 
 for  $ \dot m \le $ 1, $R_{\rm s}/r_{\rm g} \sim $ 11 --12. 
 The transition of the former shock position to the latter one occurs in the 
  range of $\dot m = 1$ -- $10^2$. 
 The shock positions  in the very high accretion rates 
  agree with the theoretical adiabatic solution. 
 In these cases, the gas is optically thick and radiation-pressure dominant, 
 and the radiative cooling term balances the radiative heating one, that is, 
  the radiative energy source $\Lambda \propto (S_* - E_{\rm 0}) \sim$ 0.  
  Accordingly the flow behaves as the adiabatic flow.
  However, in the low accretion rates where the gas is optically thin,
  $ S_* (=a T^4) \gg E_0 $ and  
  the balance of cooling and heating is not established anymore.
  The source function $\Lambda$ works as the cooling source.
   At the Rankine--Hugoniot relations of a standing radiative shock in the high 
  accretion rate, the pressure balance is supported by the sum of dominant radiation 
  pressure and ram pressure.
  When the ambient density is taken to be much lower than that in the high 
  accretion rate, the upstream pressure just before the shock is much smaller 
  because of the lower temperature in the pre-shock region.
   Therefore, to set up a new pressure balance condition at the shock,
   the shock must shift outward  as far as the same injection parameters 
   are concerened.
  This is the reason why the decreasing input accretion rate leads to
  the increasing shock position on the equatorial plane. 
   If the ambient density is taken to be too low, there exists no longer 
   shock wave under the same injection parameters, because the parameters which 
   are originally derived from the adiabatic equations could not match with 
   the optically thin flow.
   
   The QPO-behaviors of the luminosity were attributed to the shock oscillations. 
   We furthermore consider that the shock oscillations are driven by the centrifugal
   barrier and the radiative cooling in the shocked region \citep{b19}.  
   When the gas is fully optically thick, the cooling 
   effect is negligible. 
   In the transonic accretion problems around the black hole, it is well known
   that the transonic flows have generally multi-shock wave solutions, such as 
   the inner and outer shock solutions. The shock waves obtained in this
   paper correspond to the outer shock wave solution.
    From the instability analyses of accretion flow with a standing shock wave, 
   it has been found that the outer shock wave both for isothermal and adiabatic 
   flows are dynamically stable \citep{b23,b24}. 
   On the other hand, when the flow is in an optically thin state, $ S_* \gg E_0 $ and
   the source function $\Lambda \propto \kappa \rho T^4$. 
   As the shock is perturbed and propagates outward, it heats the post-shock flow
   to a higher temperature since the relative velocity between the shock and the 
   incoming flow becomes higher. As a result, the radiative cooling in the post-shock
   region grows up and the outward motion of the shock stops when the flow is
   sufficiently cooled down. The shock eventually moves towards the 
   black hole due to the reduced pressure behind the shock.
   This time, the relative velocity between the shock and the post-shock flow
   decreases and the post-shock temperature drops.  The lower pressure in the 
   post-shock region is unable to balance the preshock ram pressure and 
   the shock collapse continues till the centrifugal barrier is sufficient to hold
   the flow. The shock then bounces outward and this process is repeated as the shock 
   oscillations.
   In order to have an oscillatory behavior, the post-shock region must be able to cool 
   in a cooling timescale  comparable to the advection timescale at the shock. 
   Thus far, the shock oscillation period is considered as the advection timescale at 
   the shock position $R_{\rm s}$ and the larger $R_{\rm s}$ leads to the larger
  oscillation timescale and the smaller QPO-frequency.
  From the power spectra in Figs.~5 and 10, we find that the
  QPO-frequencies drift to the larger frequencies with increasing $\dot m$.

     \begin{figure}
     \includegraphics[width=86mm,height=66mm]{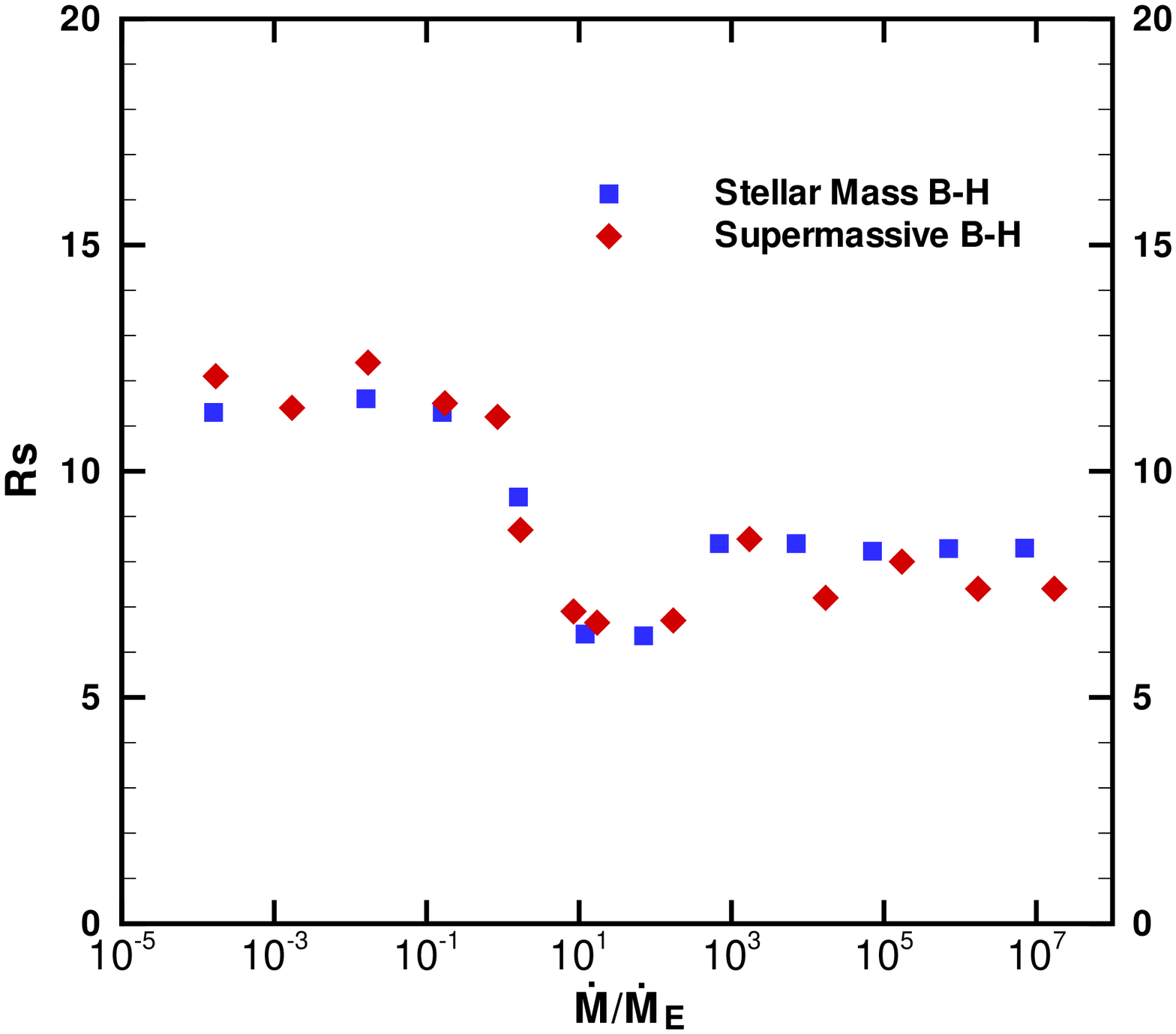}
      \caption{Shock position $R_{\rm s}$ on the equatorial plane versus $\dot m$
       for cases of the stellar-mass  and the supermassive black holes}  
      \label{fig11}
      \end{figure}

 Fig.~12 shows the normalized luminosity $L/L_{\rm E}$
 versus $\dot m$ in the  wide range of
 $10^{-4}$ -- $10^7$. The $L/L_{\rm E} - \dot m$ relations are almost 
 identical both for the stellar-mass and the supermassive black holes. 
 When  $\dot M$ is low, the luminosity increases in proportion to the accretion rate.
 However, when $\dot M$  exceeds greatly  the Eddington critical rate $\dot M_{\rm E}$, 
 the luminosity is insensitive to $\dot m$ and is kept 
 constantly around a maximum luminosity of $\sim 3 L_{\rm E}$. We have here approximately 
 
  \begin{equation}
   {\rm log}{ L \over L_{\rm E} }=\cases {{\rm log}\; \dot m -1.2,
                    \;\; \;\; {\rm for}\;  \dot m \le 10^{-2} \cr
   -0.15 \;({\rm log} \;\dot m)^2 +0.74\;{\rm log}\; \dot m - 0.81, 
      \;\; \cr 
  \;\;\;\;\;\;\;\;\;\; \;\;\;\;\;\;\;\;\;\;\;\;\;{\rm for} \; 
                            10^{-2} \le \dot m \le 10^2, \cr
   0.48, \;\;\;\;\;\;\;\;\;\;\;\; \;\;\;{\rm for}\; \dot m \gg 10^2. \cr}
  \end{equation}
  
      \begin{figure}
      \includegraphics[width=86mm,height=66mm]{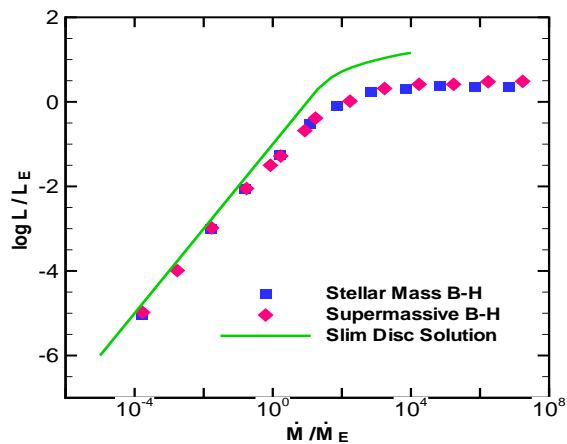}
      \caption{ Normalized luminosity $L/L_{\rm E}$ versus $\dot m$ 
      for cases of the stellar-mass  and the supermassive  black holes. 
      The line shows  the slim disc solution by Watarai 
       {\it et al}. (2000).}        
      \label{fig12}
      \end{figure}

      \begin{figure}
      \includegraphics[width=86mm,height=66mm]{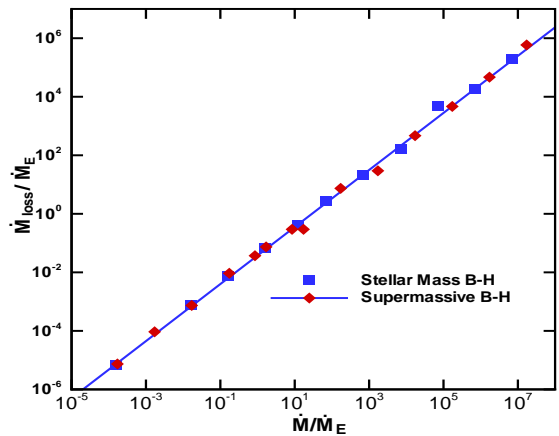}
      \caption{Mass-outflow rate $\dot M_{\rm loss}/\dot M_{\rm E}$ versus $\dot m$ 
      for cases of the stellar-mass and the supermassive black holes. }  
      \label{fig13}
      \end{figure}

 \citet{b30} studied the slim-accretion disc model shining at the Eddington 
 luminosity and numerically derived  an $L/L_{\rm E}$ - $\dot m$ relation. 
 The result is also plotted  in Fig.~12. 
 Their $L/L_{\rm E}$ - $\dot m$ relation agrees well with ours 
 in the low accretion rates but the luminosities in the high accretion rates are rather
 larger than ours. The result for the low accretion rates
 shows $L = 0.063 \dot M c^2$, that is, the conversion efficiency $\eta$ 
 of gravitational energy
 to radiation  is $\sim 1/16$. The existence of the maximum luminosity $\sim 3 L_{\rm E}$
 is interesting from  the ultra luminous X-ray sources point of view.
 The normalized mass-outflow rate $\dot m_{\rm loss} (=\dot M_{\rm loss}/\dot M_{\rm E})$ 
 versus $\dot m$  is given in Fig~13.
 The mass-outflow rate increases  in proportion to $\dot m$ and we have 
 
 \begin{equation}
  \dot M_{\rm loss} =  0.04\; \dot M.
 \end{equation}
 The mass-outflow rate $\dot M_{\rm loss}$ amounts to about a few percent of 
 the input accretion rate $\dot M$.

\section{Concluding Remarks}
 We  examined numerically 2D inviscid 
 transonic flows around the stellar-mass and the supermassive black holes, 
 while taking account of the cooling and heating of  gas and 
 radiation transport.  In these accretion flows, 
 the centrifugally supported shock waves are formed around the black holes. 
 The shock waves are unstable with time and the resultant 
 luminosities show the QPOs with modulations of a factor
  of 2 -- 3. 
 The shock positions are weakly dependent on the accretion rate over a wide range
 of $\dot m$ and are in the range of 7 -- 12 $r_{\rm g}$ on the equatorial plane.  
 In the cases of much higher accretion rates ($\dot m \gg 1$), 
 the shock position agrees with the theoretical adiabatic solution .
 On the other hand, the much lower accretion rate leads to the more outward shock position
 and accordingly the smaller QPO-frequency. 
 As results, we have
 the QPO frequency $\nu_{\rm qpo}$ of a few to 10 Hz and $3\times 10^{-5}$
  -- $3\times 10^{-4}$ Hz, that is, the period $P_{\rm qpo}$ of  0.1 -- several seconds and
  1 -- $10$ hours for the stellar-mass and the supermassive black holes,
  respectively.

 \citet{b3} examined numerically  the shock oscillation 
 model around the black holes with masses of $10 M_{\rm \odot}$ and $10^8 M_{\rm \odot}$
 and obtained the QPO-frequencies of $\sim$ 8 -- 12 Hz and  $10^{-7}$ -- $10^{-6}$
 Hz for the stellar-mass and the supermassive black holes, respectively.
 We plot the QPO period-mass relation in Fig.~14, together with their results.
 Here we have approximately
 
 \begin{equation}
     P_{\rm qpo} =  0.016 ({M_* \over M_{\rm \odot}}) \;\;{\rm sec}. 
 \end{equation}
 This suggests the QPO-periods expected from the shock oscilation model around 
 black holes in various mass range.
 
 The normalized luminosity and the mass-outflow rate versus $\dot m$ 
 have common properties independently on the black hole masses. 
 When  $\dot m$ is low, the luminosity increases in proportion to the accretion rate.
 However, when $\dot M$  exceeds greatly  the Eddington critical rate $\dot M_{\rm E}$, 
 the luminosity is insensitive to the accretion rate and is kept 
 constantly around 3 $L_{\rm E}$. On the other hand, the mass-outflow rate 
 $\dot M_{\rm loss}$ increases in proportion to $\dot M$  and it amounts to
 a few percent of the input mass-flow rate both for the 
 stellar-mass and the supermassive black holes.

      \begin{figure}
      \includegraphics[width=86mm,height=66mm]{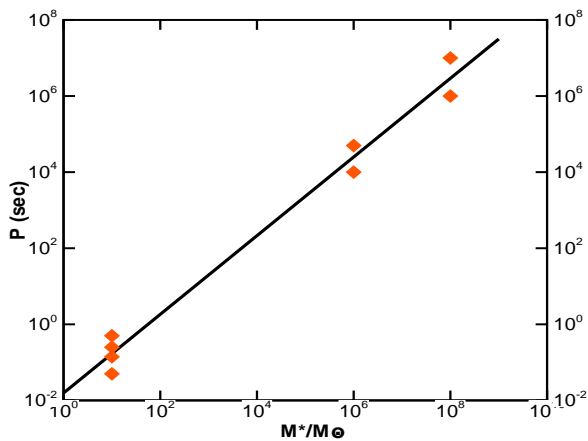}
      \caption{QPO-period $P_{\rm qpo}$ versus black hole mass $M_*/M_{\rm \odot}$ due to 
       the  shock oscillation model.}
      \label{fig14}
      \end{figure}

  The shock oscillation model works only for the accretion flows
  within a limited parameter space with adequate input parameters for 
  the inviscid flows. Up to now, only the transonic solutions of the accretion flows 
  with total positive energy have been discussed and simulated.
  However, recent work of the sub-Keplerian flows with negative total
  energy  enlarges the parameter space for which the steady shocks 
  are exhibited in the rotating accretion flows around the black holes \citep{b17}. 
  Even in the viscous accretion discs, the shock oscillation models are still valid 
  when the 
  viscosity parameter, $\alpha$, is less than a critical value \citep{b19,b15}. 
  Of course, if the viscosity is too high, the standing shock disappears.

 The centrifugally supported shocks around black holes and the shock oscillation models 
 have been applied to the QPO phenomena in the black hole 
 candidate GRS 1915+105 \citep{b2,b4,b28} in comparison with the observational data.
 The Galactic X-ray transient source GRS 1915+105  exhibits 
 various types of QPOs: 
 (1) the low-frequency QPO ($\nu_{\rm L} \sim $
   0.001 -- 0.01Hz), (2) the intermediate frequency QPO ($\nu_{\rm I} \sim$
    1 -- 10Hz), and (3) the high frequency QPO ($\nu_{\rm H} \sim$
     67 Hz) \citep{b22}.
  The QPO-frequencies of a few to 10 Hz due to the shock oscillation model
  in the stellar-mass 
  black hole may  be representative of the intermediate frequency QPO observed 
  in GRS 1915+105.

 The shock oscillation model of the supermassive black hole may be applicable to 
 Active Galactic Nuclei, because they are powered by accretion onto a central
 black hole. Therefore, it may be plausible that QPOs 
 should be observed in AGNs as well. The QPO-frequency due to the shock 
 oscillation model depends on the mass of the supermassive black hole.
 Recent data of Seyfert galaxies by the EUVE and the Chandra show 
 some periodicities or QPOs in the range of a tenth of hours to a month 
 \citep{b12,b21} and may suggest the relevance to the shock oscillation model.
 Further observations of the black hole candidates will be needed to confirm the
 shock oscillation model for QPOs.

\label{lastpage}


\begin{thebibliography}{99}

\bibitem[\protect\citeauthoryear{Chakrabarti}{1989}]{b1}
  Chakrabarti, S. K.,  1989, ApJ, 347, 365


\bibitem[\protect\citeauthoryear{Chakrabarti}{1999}]{b2}
  Chakrabarti, S. K.,  1999, A\&A, 351, 185

\bibitem[\protect\citeauthoryear{Chakrabarti et al.}{2004}]{b3}
 Chakrabarti, S. K., Acharyya., Molteni, D., 2004, A\&A, 421, 1

\bibitem[\protect\citeauthoryear{Chakrabarti \& Manickam}{2000}]{b4}
  Chakrabarti, S. K.,  Manickam, S. G.,  2000, ApJ, 531, L41


\bibitem[\protect\citeauthoryear{Chakrabarti \& Molteni}{1993}]{b5}
  Chakrabarti, S. K.,  Molteni, D.,  1993, ApJ, 417, 671


\bibitem[\protect\citeauthoryear{Das}{2002}]{b6}
  Das,  T. K.,  2002, ApJ, 577, 880

\bibitem[\protect\citeauthoryear{Das}{2003}]{b7}
  Das, T. K.,  2003, ApJ, 588, L89

\bibitem[\protect\citeauthoryear{Das et al.}{2001}]{b8}
  Das, S., Chattopadhyay, I., Chakrabarti, S. K., 2001, ApJ, 557, 983
  

\bibitem[\protect\citeauthoryear{Das et al.}{2003a}]{b9}
  Das, T. K., Pendharkar, J. K.,  Mitra, S., 2003a, ApJ, 592, 1078

\bibitem[\protect\citeauthoryear{Das et al.}{2003b}]{b10}
  Das, T. K., Rao, A. R., Vadawale, S. V., 2003b, MNRAS, 343, 443

\bibitem[\protect\citeauthoryear{Fukue}{1987}]{b11}
  Fukue, J., 1987, PASJ, 39, 309

\bibitem[\protect\citeauthoryear{Halpern et al.}{2003}]{b12}
  Halpern, J. M., Leighly, K. M..,  Marshall, H. L..,  2003, ApJ, 585, 665


\bibitem[\protect\citeauthoryear{Kato et al.}{1998}]{b13}
  Kato, S., Fukue, J., Mineshige, S. ,1998, Black Hole Accretion Disks
  (Kyoto: Kyoto University Press)
 
 \bibitem[\protect\citeauthoryear{Kley}{1989}]{b14}
  Kley, W.,  1989, A\&A, 208, 98
  
  \bibitem[\protect\citeauthoryear{Lanzafame et al.}{1998}] {b15}
  Lanzafame, G., Molteni, D.,  Chakrabarti, S. K.,  1998, MNRAS, 299, 799
  
\bibitem[\protect\citeauthoryear{Levermore \& Pomraning}{1981}]{b16}
  Levermore, C. D.,  Pomraning, G. C.,  1981, ApJ, 248, 321

\bibitem[\protect\citeauthoryear{Molteni et al.}{2006}]{b17}
  Molteni, D., Gerardi, G.,  Teresi, V.,  2006, MNRAS, 365, 1405


\bibitem[\protect\citeauthoryear{Molteni et al.}{1994}]{b18}
  Molteni, D., Lanzafame, G.,  Chakrabarti, S. K.,  1994, ApJ, 425, 161


\bibitem[\protect\citeauthoryear{Molteni et al.}{1996}]{b19}
  Molteni, D., Sponholz, H.,  Chakrabati, S. K.,  1996, ApJ, 457, 805
  

\bibitem[\protect\citeauthoryear{Molteni et al.}{1999}]{b20}
  Molteni, D., Toth, G., Kuznetsov, O. A., 1999, ApJ, 516, 411

\bibitem[\protect\citeauthoryear{Moran et al}{2005}]{b21}
  Moran, E. C., Eracleous, M.,  Leighly, K. M.,  Chartas, G., 
  Filippenko, A. V., Ho, L. C., Blanco, P. R., 2005, AJ, 129, 2108

\bibitem[\protect\citeauthoryear{Morgan}{1997}]{b22}
  Morgan, E. H., Remillard, R. A.,  Greiner, J.,  1997, ApJ, 482, 993

 
 \bibitem[\protect\citeauthoryear{Nakayama}{1994}]{b23}
  Nakayama K.,  1994, MNRAS, 270, 871
 
 \bibitem[\protect\citeauthoryear{Nobuta \& Hanawa}{1994}]{b24}
  Nobuta K., Hanawa T.,  1994, PASJ, 46, 257
 

\bibitem[\protect\citeauthoryear{Okuda et al.}{1997}]{b25}
  Okuda, T.,  Fujita, M.,  Sakashita, S.,  1997, PASJ, 49, 679
 
  
\bibitem[\protect\citeauthoryear{Okuda et al.}{2004}]{b26}
  Okuda, T., Teresi, V., Toscano, E.,  Molteni, D., 2004, PASJ, 56, 547

\bibitem[\protect\citeauthoryear{Paczy\'{n}sky \& Wiita}{1980}]{b27}
  Paczy\'{n}sky, B.,  Wiita, P. J.,  1980, A\&A, 88, 23

\bibitem[\protect\citeauthoryear{Rao et al.}{2000}]{b28}
 Rao, A. R., Naik, S., Vadawale, S. V.,  Chakrabarti, S. K.,  2000, A\&A, 360, L25
 
\bibitem[\protect\citeauthoryear{Ryu et al.}{1997}]{b29}
  Ryu, D., Chakrabarti, S. K.,  Molteni, D.,  1997, ApJ, 474, 378

\bibitem[\protect\citeauthoryear{Watarai et al.}{2000}]{b30}
  Watarai, K., Fukue, J., Takeuchi, M,  Mineshige, S.,  2000, PASJ, 52, 133
 
\end{thebibliography}
\end{document}